\documentclass[runningheads,a4paper]{llncs}

\usepackage{times}
\usepackage{graphicx}
\usepackage{latexsym}
\usepackage{amssymb}
\usepackage{xcolor}
\usepackage[hyphens]{url}
\usepackage{float}
\usepackage[procnames]{listings}
\usepackage{tikz}
\usepackage{pgf-umlcd}
\usepackage{array,multirow,makecell}
\setcellgapes{1pt}
\makegapedcells
\usepackage{colortbl}
\definecolor{Gray}{gray}{0.85}

\usepackage{subfig}

\definecolor{mygray}{rgb}{0.4,0.4,0.4}
\definecolor{mydarkblue}{rgb}{0.0,0.0,0.4}
\definecolor{mycyan}{rgb}{0.0,0.4,0.4}
\definecolor{mygreen}{rgb}{0.0,0.5,0.0}
\lstdefinelanguage{turtle}
{
  columns=fullflexible,
  morekeywords={@prefix},
  keywordstyle=\color{mygray},
  morecomment=[l]{\#},
  alsoletter={-?}, % allowed in names
  moredelim=*[s][\color{mygreen}]{_:bn}{\ },
  morecomment=[s][\color{mydarkblue}]{<}{>},
  breaklines=true,
  aboveskip=6pt,
  numberstyle=\color{darkblue},
  showstringspaces=false,
  framexleftmargin=0mm,
  frame=shadowbox, 
  framextopmargin=2pt,
  framexbottommargin=2pt,
  rulesepcolor=\color{gray},
  rulesep=0.5pt,
  captionpos=b,
  morestring=[b][\color{black}]\",
}

\definecolor{olivegreen}{rgb}{0.2,0.8,0.5}
\definecolor{gray}{rgb}{0.5,0.5,0.5}
\lstdefinelanguage{ttl}{
sensitive=true,
morecomment=[l][\color{gray}]{@},
morecomment=[l][\color{olivegreen}]{\#},
basicstyle=\ttfamily,
morestring=[b][\color{blue}]\",
}

\newcolumntype{R}[1]{>{\raggedleft\arraybackslash }b{#1}}
\newcolumntype{L}[1]{>{\raggedright\arraybackslash }b{#1}}
\newcolumntype{C}[1]{>{\centering\arraybackslash }b{#1}}

\urldef{\mailsa}\path|{Nathalie-F.Abadie, Benedicte.Bucher, Abdelfettah.Feliachi}@ign.fr|

\title{GeomRDF: A Geodata Converter with a Fine-Grained Structured Representation of Geometry in the Web}%
\titlerunning{A Fine-Grained Structured Representation of Geometry in the Web}
\author{Fay\c cal Hamdi\inst{1}  \and Nathalie Abadie\inst{2} \and B\'en\'edicte Bucher \inst{2} \and Abdelfettah Feliachi \inst{2}
}
\institute{
CEDRIC, CNAM, F-75141 Paris Cedex 03, France \\
\email{Faycal.Hamdi@cnam.fr}
\and
Universit\'e Paris-Est, IGN/SRIG, COGIT, 73 Avenue de Paris, 94160 Saint-Mand\'e, France
\mailsa
}

\begin{document}

\maketitle

\begin{abstract}
In recent years, with the advent of the web of data, a growing number of national mapping agencies tend to publish their geospatial data as Linked Data. However, differences between traditional GIS data models and Linked Data model can make the publication process more complicated. Besides, it may require, to be done, the setting of several parameters and some expertise in the semantic web technologies. In addition, the use of standards like GeoSPARQL (or ad hoc predicates) is mandatory to perform spatial queries on published geospatial data. In this paper, we present GeomRDF, a tool that helps users to convert spatial data from traditional GIS formats to RDF model easily. It generates geometries represented as GeoSPARQL WKT literal but also as structured geometries that can be exploited by using only the RDF query language, SPARQL. GeomRDF was implemented as a module in the RDF publication platform Datalift. A validation of GeomRDF has been realized against the French administrative units dataset (provided by IGN France). %\keywords{Linked Data, Geospatial Data}
\end{abstract}

\section{Introduction}

Over the last decade, efforts to share geospatial data on the Web have mainly focused on the development and the standardization by ISO TC 211 and the  Open Geospatial Consortium (OGC) of domain-specific good practices and tools called Spatial Data Infrastructures (SDIs) \cite{craglia08a}. These good practices have been particularly confirmed by the adoption by the European Parliament of the INSPIRE Directive establishing an infrastructure for spatial information in the Community \cite{inspire}.  At the same time, generic good practices for data sharing on the Web called Linked Data have been developed and supported by the W3C \cite{bernerslee06}. Over the last years, many datasets have been published according to the Linked Data principles, so that these good practices have spread and are now being adopted for geospatial data.  For example, Geo.LinkedData.es \cite{geolinkeddataes}, is an initiative to enrich the Web of data with Spanish geospatial data. Besides, LinkedGeoData \cite{linkedgeodata}, is an effort to add information collected by the OpenStreetMap project to the Web of data. Moreover, the Ordnance Survey Linked Data \cite{ordnancesurvey} is an initiative of the Great Britain's national mapping agency to publish a number of its products as Linked Data.\\

Although these initiatives are promising, the amount of geospatial data published as Linked Data remains limited compared to available geospatial data published through other means like Spatial Data Infrastructures or Open Data platforms. This could be partly due to the fact that converting geospatial data from their traditional GIS formats to fully interconnected RDF data raises a domain-specific difficulty: converting into RDF the geometries used by geographic information standards for representing the location and the shape of real world geographic entities. This difficulty occurs at the first step of the publishing process. To be overcome, it requires not only geolocation vocabularies, but also specific tools that automatise the conversion of existing geospatial data into Linked Data \cite{zimmermann2010a}.\\

To deal with these issues, we present the GeomRDF tool which offers to geospatial data publishers the possibility to publish (in an easy-way) their data, and to users the possibility to query these geospatial data, using semantic web technologies. Most importantly, the GeomRDF tool bundles a fine-grained representation of geometry based on a vocabulary \cite{geom} that re-uses and extends the existing geographic standards and vocabularies (GeoSPARQL \cite{geosparql2}, NeoGeo \cite{neogeo}). This representation enables automated agents to reason over geometries. GeomRDF is implemented as a module\footnote{The current release of the Datalift Platform (available at: \url{https://gforge.inria.fr/scm/?group_id=2935}) includes the experimental version of GeomRDF (that does not contain the geometry structuring process). The final version of GeomRDF code source will be published with the next Datalift release.} in the Datalift Platform \cite{datalift} in order to provide a complete path from geospatial data (ESRI Shapefile,DBMS and GML) to fully interlinked, identified, and qualified linked data.\\

The paper is organized as follows. In the next section, we present  some related works. In section 3 we present in detail the main components of the GeomRDF tools, together with the supported geospatial formats and external libraries. In section 4 we present a test of these tools on a dataset that represents French administrative units, provided by IGN France. Finally we conclude and give some perspectives in Section 5.

\section{Related Works}

Real world phenomenon are represented in geographic vector databases by geographic features, described by thematic properties and a geometry. Thus, converting geospatial data from their original standard to RDF implies representing not only their thematic properties - which can be handled like any other data, but also their associated geometries in RDF. Many knowledge extraction and data conversion tools have been proposed to generate RDF data from unstructured or structured data sources \cite{unbehauen12a} \cite{bizer2006a} \cite{d2rqlang} \cite{auer09a} \cite{sponger} \cite{poolparty} \cite{lebo11a}, but just a few vocabularies and their associated conversion tools have been implemented to address the issue of converting easily geospatial data into the RDF model.\\
Geometry2RDF \cite{geo2rdf} is a tool, developed  in the context of Geo.LinkedData.es \cite{geolinkeddataes}, that enables geospatial data conversion into RDF. The conversion process takes as input different geospatial formats (ESRI Shapefile, GML and geospatial DBMS) and generates RDF triples according to the NeoGeo vocabulary \cite{neogeo}. Consistently with NeoGeo vocabulary, geometries associated to geographic features are therefore represented in RDF as structured geometries. NeoGeo vocabulary defines geometries following OGC's GML Simple Features Profile. This vocabulary directly reuses the class Point of the W3C Geo Vocabulary \cite{w3cgeo}, and its related properties. As a consequence, geometries represented with NeoGeo vocabulary can be defined only with WGS84 coordinates.

TripleGeo \cite{patroumpas14} is an extension of Geometry2RDF that takes into account, in the transformation process, the GeoSPARQL vocabulary. This vocabulary for representing RDF geospatial data has been defined as part of the OGC standard GeoSPARQL, which also provides an extension to the SPARQL query language  for processing RDF geospatial data. This vocabulary defines two RDFS datatypes , namely \url{http://www.opengis.net/ont/geosparql#wktLiteral} and \url{http://www.opengis.net/ont/geosparql#gmlLiteral}, for representing geometries respectively as WKT serialization as defined by the Simple Feature standard \cite{iso19125} and as GML serialization as defined by GML standard \cite{gml}. Both serializations imply expliciting for each geometry the coordinate reference system (CRS) used. For that purpose, OGC maintains a set of CRS URIs for identifying the most common CRSs. In addition, TripleGeo provides an on-the-fly functionality for transforming geometries coordinates from one given coordinate reference system to another.

shp2GeoSPARQL \cite{velasquez14a} is also an extension of Geometry2RDF which transforms geometrical information from geospatial datasets to RDF. shp2GeoSPARQL parses ESRI Shapefile in order to retrieve the geometries of features, and generates a RDF representation of geometries consistent with the GeoSPARQL geometry vocabulary.

These three solutions have their advantages and drawbacks. On the one hand, the tools based on GeoSPARQL vocabulary have the advantage of  allowing geometries defined in any CRS. However, there are very few triplestores that implement this standard yet \cite{battle12a}, so that GeoSPARQL spatial operators and functions can not be used commonly with the available linked geospatial data. On the other hand, the tool based on NeoGeo vocabulary has the advantage of providing structured geometries that can be handled with regular SPARQL queries. But geometries coordinates are restricted to WGS84 CRS. In order to overcome these limitations, we propose the GeomRDF tool. It is based on a vocabulary that reuses and extends GeoSPARQL simple feature vocabulary and NeoGeo so that geometries can be defined in any CRS and represented both as structured geometries that can be handled with regular SPARQL queries and as wktLiteral datatypes that are compliant with GeoSPARQL standard.
%\vspace{-0.26cm}
\section{GeomRDF Components}

\begin{figure}
  \centering
  \includegraphics[width=11cm]{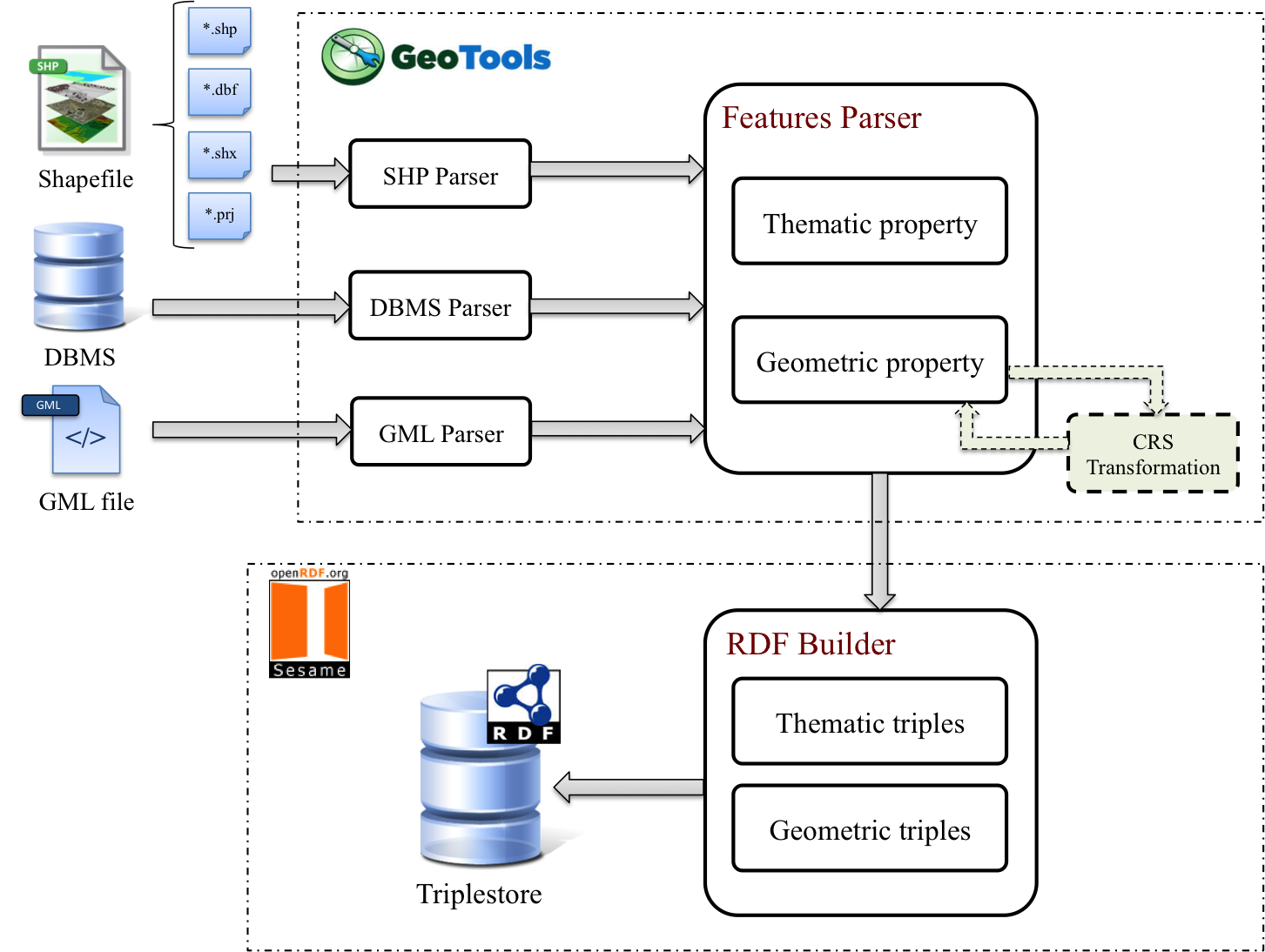}
  \caption{GeomRDF Components}
   \label{GeomRDComp}
\end{figure}

In this section we describe the different components of GeomRDF (cf. Fig.~\ref{GeomRDComp}) and the process that transforms a geospatial dataset from traditional GIS formats to RDF. GeomRDF includes three components: the input parser, the features parser and the RDF builder. We outline in the following sections these components.

\subsection{The Input Parser:} GeomRDF takes as input ESRI Shapefile \cite{shapefile}, Geospatial DBMS or GML \cite{gml}. This component extracts from each file/database all the features and their descriptions (e.g., schema, types, CRS (Coordinate Reference System). We mean by "feature", the "Simple Feature" defined by the Open Geospatial Consortium (OGC) \cite{simplef} and the International Organization for Standardization (ISO) standard ISO 19125  to have both spatial and non-spatial attributes. Spatial attributes are geometry valued, and simple features are based on 2D geometry with linear interpolation between vertices (e.g., Multipolygones, Polygons, .etc).
\subsubsection{ESRI Shapefile:} This format is a part of the GIS technology offered by ESRI. Spatial data format defined by ESRI stores non-topological geometry and attribute information for the spatial features in a dataset.\\
Regarding this format, four files are processed: the three mandatory dBASE file (.dbf), index file (.shx) and main file (.shp), plus the metadata file (.prj) that describes the CRS used by the dataset (In the case where this file is missing, the default considered CRS is WGS84). The schema and the type of the thematic and geometric attribute are extracted from the main and dBase file.
\subsubsection{Geospatial DBMS:}  It is a kind of database that is optimized to store and query geospatial data. GeomRDF can handle: IBM DB2,  H2, MySQL, Oracle Spatial, PostGIS (a spatial extension to PostgreSQL), SpatiaLite (a spatial extension to SQLite), Microsoft SQL Server and Teradata.
\subsubsection{GML:} OGC standard encoding specification for geodata in XML that enables the storage, transport, processing, and transformation of geographic information.\\

These three format of storage are well known in the geographic information science field and are used by the majority of mapping agencies.
 
\subsection{Feature Parser:} 
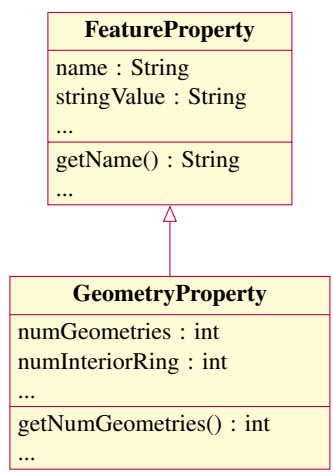
\begin{figure}[!htb]
\centering
\begin{tikzpicture}
%\tikzstyle{every node}=[font=\scriptsize]
  \begin{class}[text width=3cm]{FeatureProperty}{0,0}
    \attribute{name : String}
    \attribute{stringValue : String}
     \attribute{...}
     \operation{getName() : String}
     \operation{...}
  \end{class}
  \begin{class}[text width=4cm]{GeometryProperty}{0,-3.5}
    \inherit{FeatureProperty}
    \attribute{numGeometries : int}
    \attribute{numInteriorRing : int}
     \attribute{...}
    \operation{getNumGeometries() : int}
    \operation{...}
  \end{class}
\end{tikzpicture}
\caption{Feature Properties Class Diagram}
\label{fig:class}
\end{figure}

This component iterates on the features extracted during the first step and classifies their properties depending on their type, either thematic or geometric. For one feature the parser tests all the properties and, for each property, stores information like its name, its value and its type. When the property is geometric, other information concerning this property (e.g., the number of geometries contained by a Multipolygon) will be stored. The Fig.~\ref{fig:class} shows the two classes designed to store feature properties.
%\vspace{-3.5pt}

To illustrate how features are processed, we take as example the department of "Paris" defined in GEOFLA dataset on French administrative units produced by IGN France and stored in ESRI Shapefile format. We detail thereafter the different steps for storing thematic and geometric properties of this feature.

\subsubsection{Thematic properties}
All the thematic properties that describe the department of "Paris" are stored as a "FeatureProperty" structure (cf. fig~\ref{fig:class}). These properties include, for instance, the department code (CODE\_DEPT) the department name (NOM\_DEPT), the region code (CODE\_REG) and the region name (NOM\_REGION) (the other properties are detailed in the specifications of this dataset\footnote{\url{http://professionnels. ign.fr/sites/default/files/DC_GEOFLA_1-1.pdf}}).\\
The following table shows a sample of the stored thematic properties, their values and their types:
\begin{table}[H]
\centering
\begin{tabular}{ |L{1.3cm}|L{1.8cm}|L{1.8cm}|C{2.5cm}|L{1.8cm}|L{2.4cm}| }
   \hline
   \rowcolor{Gray}
   Property & CODE\_DEPT & NOM\_DEPT & ...  & CODE\_REG & NOM\_REGION\\
   \hline
   Value & 75 & PARIS & ... & 11 & ILE DE FRANCE\\
    \hline
   Type & \texttt{String} & \texttt{String} & ... & \texttt{String} & \texttt{String} \\
    \hline 
\end{tabular}
\vspace{2pt}
\caption{The thematic properties of the department of "Paris"}
\label{tab1}
\end{table}
\vspace{-1.5cm}
\subsubsection{Geometric properties}
The geometry associated to the department of "Paris" is stored as a sequence of coordinates defining the points that are part of the boundary of the department. Because of the length of this geometry, we represent only some points:\\
%\small\texttt{MULTIPOLYGON (((2.416339717283465 48.84923827341306, 2.4159737754767168 48.84662836540369, .... , 2.416339717283465 48.84923827341306)))"}

{\noindent{\small\texttt{MULTIPOLYGON(((2.41633 48.84923, 2.41597 48.84662, .... , 2.41633 48.84923)))}}\\

This geometry is stored as "GeometryProperty" structure (cf. fig~\ref{fig:class}). In this example, the geometry of the department of "Paris" is represented by a Multipolygon. Indeed, this type of geometry is a good example because it includes the majority of the other types of geometry handled by GeomRDF.
In this case, the parser stores at first all the Polygons that compose the original Multipolygon. Then, it stores the exterior and the eventual interior LinearRings that compose each polygon, as well as all the points included in each LinearRing. Finally the coordinates of each point are stored.\\

At this stage, GeomRDF provides a very useful functionality which consists in transforming these coordinates from their original CRS (Coordinate Reference System) to another one. This transformation is performed on-the-fly; before storing the geometry, the parser check whether there is a need or not to convert the CRS.

\subsection{RDF Builder:}
The RDF Builder generates, from the parsed features, a collection of RDF triples expressed as subjects, predicates, and objects. As in the Feature Parser component, thematic and geometric properties are treated separately. We illustrate in the following, the process of generating RDF triples from the feature (the department of "Paris"), parsed in the above example.
\vspace{-4.4pt}
\subsubsection{Thematic properties}
Consider the following base URI \lstinline[basicstyle=\small\ttfamily,breaklines=true,language=turtle]{data:<http://data.ign.fr/id/geofla/departement/>}. The value of a property is captured as text with the appropriate RDF literal type. For instance the value of the NOM\_REGION will be represented by the literal \lstinline[basicstyle=\small\ttfamily,breaklines=true,language=turtle]{"ILE DE FRANCE"^^xsd:string}.\\
By default, GeomRDF generates predicates by reusing properties names. They can be replaced afterwards by predicates from the dataset vocabulary. This task of matching and replacing default predicates with well defined ontology predicates is handled within another module of the Datalift platform. Listing~\ref{ttl1} shows an example of RDF triples representing thematic properties.

\begin{lstlisting}[basicstyle=\scriptsize\ttfamily,caption=A sample of RDF thematic triples of the department of "Paris",label=ttl1,language=turtle]
@prefix rdf: <http://www.w3.org/1999/02/22-rdf-syntax-ns#>
@prefix geofla:<http://data.ign.fr/def/geofla#>

<http://data.ign.fr/id/geofla/departement/75> a geofla:Departement .
<http://data.ign.fr/id/geofla/departement/75> rdfs:label "PARIS"@fr .
<http://data.ign.fr/id/geofla/departement/75> geofla:codeDpt "75"^^xsd:string .
\end{lstlisting}
\vspace{-0.58cm}
\subsubsection{Geometric properties}
The generation of the geometric RDF triples is guided by a vocabulary \cite{geom} that extends GeoSPARQL standard and gives a very precise description of the different types of geometry. Based on this vocabulary, GeomRDF generates, for each geometric property stored by the Feature parser, the appropriate RDF triples that represent the geometry and its CRS. Furthermore, as the the vocabulary of geometries was built in OWL language, a semantic reasoner can check the consistency and infer new axioms against generated triples. Each instance of a geometric class is linked to the URL identifying the CRS used for its coordinates thanks to the \texttt{geom:crs} predicate. To well detail the functionality of this component we take as example the process of generating RDF triples from the geometric property of the department of "Paris".\\

In order to structure a Multipolygon, we follow the axioms defined in the vocabulary of geometries.  We outline thereafter, the axioms used to define a Multipolygon and some samples of the generated RDF triples.\\

\textbf{Axiom 1 (Multipolygon)} \textit{ An instance of the class \texttt{\bfseries geom:MultiPolygon} is associated to one or more other instances of the class \texttt{\bfseries geom:Polygon} via the property\break \texttt{\bfseries geom:polygonMember}.}
%The RDF triples generated for the Multipolygon of of the department of "Paris" are:
\begin{lstlisting}[basicstyle=\scriptsize\ttfamily,caption=A sample of RDF triples representing a Multipolygon,language=turtle]
@prefix geom:<http://data.ign.fr/def/geometrie#>

<http://data.ign.fr/id/geofla/departement/75> a geom:MultiPolygon ;
geom:crs <http://data.ign.fr/id/ignf/crs/WGS84GDD> ;
geom:polygonMember  _:bn00001 ;
geom:polygonMember  _:bn00002 .
\end{lstlisting}

\textbf{Axiom 2 (Polygon)} \textit{An instance of the class \texttt{\bfseries geom:Polygon} is defined by its contour which is an instance of the class \texttt{\bfseries geom:LinearRing} associated with the instance of \texttt{\bfseries geom:Polygon} via the property \texttt{\bfseries geom:exterior}. A polygon has a single contour. It may also contain holes (0 or more), defined as instances of\break \texttt{\bfseries geom:LinearRing} associated to \texttt{\bfseries geom:Polygon} via the property\break \texttt{\bfseries geom:interior}.}
%The RDF triples representing the Polygons associated to the Multipolygon of the department of "Paris" are:
%\begin{center}
\begin{lstlisting}[basicstyle=\scriptsize\ttfamily,caption=A sample of RDF triples representing Polygons,language=turtle]
_:bn00001 a geom:Polygon ;
geom:crs <http://data.ign.fr/id/ignf/crs/WGS84GDD> ;
geom:exterior  _:bn000021 .

_:bn00002 a geom:Polygon ;
geom:crs <http://data.ign.fr/id/ignf/crs/WGS84GDD> ;
geom:exterior  _:bn000031 .
\end{lstlisting}
%\end{center}

\textbf{Axiom 3 (LinearRing)} \textit{An instance of the class \texttt{\bfseries geom:LinearRing} is associated to an instance of \texttt{\bfseries geom:PointsList}, which is a list of points, via the property \texttt{\bfseries geom:points}. This instance of PointsList must be associated to its initial point (which is also its last point) via the property \texttt{\bfseries geom:firstAndLast}.}
%The RDF triples representing the LinearRings associated to the above Polygons are:
\begin{lstlisting}[basicstyle=\scriptsize\ttfamily,caption=A sample of RDF triples representing LinearRings,language=turtle]
_:bn000021 a geom:LinearRing;
geom:crs <http://data.ign.fr/id/ignf/crs/WGS84GDD> ;
geom:points  _:bn000022 ;
geom:firstAndLast  _:bn000023 .

_:bn000031 a geom:LinearRing;
geom:crs <http://data.ign.fr/id/ignf/crs/WGS84GDD> ;
geom:points  _:bn000032 ;
geom:firstAndLast  _:bn000033 .
\end{lstlisting}

\textbf{Axiom 4 (PointsList)} \textit{A \texttt{\bfseries geom:PointsList} is a subclass of rdf:List. An instance of \texttt{\bfseries geom:PointsList} is composed of only instances of type \texttt{\bfseries geom:Point}.}
%The RDF triples representing the PointsLists are:
\begin{lstlisting}[basicstyle=\scriptsize\ttfamily,caption=A sample of RDF triples representing a PointsList,language=turtle]
_:bn000022 a geom:PointsList; geom:firstAndLast _:bn000023 ; 
rdf:rest _:bn000213 .
_:bn000213 a geom:PointsList; rdf:first _:bn000024 ; 
rdf:rest _:bn0000214 .
...
_:bn000219 a geom:PointsList; rdf:first _:bn000210 ; 
rdf:rest _:bn000022 .
\end{lstlisting}

\textbf{Axiom 5 (Point)} \textit{An instance of \texttt{\bfseries geom:Point} is associated to its coordinates via the properties \texttt{\bfseries geom:coordX} and \texttt{\bfseries geom:coordY}.}
%A sample RDF triples representing a point are:
 \begin{lstlisting}[basicstyle=\scriptsize\ttfamily,caption=A sample of RDF triples representing a Point,language=turtle]
_:bn000023 a geom:Point ;
geom:crs <http://data.ign.fr/id/ignf/crs/WGS84GDD> ;
geom:coordX "2.41633"^^xsd:double ;
geom:coordY "48.84923"^^xsd:double .
\end{lstlisting}

In addition to this structured representation of geometry coordinates,  the geometry will be associate to a WKT literal according to the GeoSPARQL standard, as the following:
\begin{lstlisting}[basicstyle=\scriptsize\ttfamily,caption=A sample of an RDF triple representing WKT literal,breaklines=true,language=turtle]
 @prefix geosparql:<http://www.opengis.net/ont/geosparql#>

<http://data.ign.fr/id/geofla/departement/75> geosparql:asWKT  "<http://data.ign.fr/id/ignf/crs/WGS84GDD> MULTIPOLYGON(((2.41633 48.84923, 2.41597 48.84662, .... , 2.41633 48.84923)))"^^<http://www.opengis.net/ont/geosparql#wktLiteral> .
\end{lstlisting}

\subsection{External Libraries}
The actual version of GeomRDF (a platform-independent version implemented in Java) depends on only two essential libraries; GeoTools \cite{geotools} and OpenRDF Sesame \cite{openrdf}. Their descriptions are listed below:
\subsubsection{GeoTools} is an open source Java library that provides methods for manipulating geospatial data. It includes many features such as a support for a wide number of formats (including ESRI Shapefile, XML (GML, KML), and geospatial databases), and a coordinate transformation service, providing coordinate reference systems descriptions and operations on coordinates such as map projections and datum transformations. The data structures used in GeoTools are based on Open Geospatial Consortium (OGC) standards.
\subsubsection{OpenRDF Sesame} is an open source Java framework for processing RDF data. Its features includes parsers, storage solutions (Triplestores), reasoning and querying, using the SPARQL query language. Sesame supports all main RDF file formats, including RDF/XML, Turtle, N-Triples, TriG and TriX.\\

In an experimental version of GeomRDF we have tested some command-line available in GDAL/OGR \cite{gdal} library. We noticed that these command-line makes our tool dependent on native library, and hence less portable. Therefore, we have decided to rule out the use of this library. 

\section{Transforming the French Administrative Units Dataset from GIS Format to RDF Model}
This section presents a test of the conversion and the publication of the French Administrative Units dataset, using the GeomRDF tool and the Datalift platform. This dataset, namely GEOFLA, is provided by the IGN France as a set of ESRI Shapefiles. It contains the description of all the administrative subdivisions (department, commune, arrondissement and canton) of France, overseas departments and Mayotte.\\
To convert and publish GEOFLA we follow the following steps: (note that all these steps are realized within the Datalift platform).
%\vspace{-0.38cm}
\subsubsection{Conversion of geospatial data to RDF} In this step we used the GeomRDF module to convert the ESRI Shapefiles to RDF. 
A transformation from the original CRS of the geometries (Lambert-93) to WGS84 is carried out. This choice is motivated by the fact that most of the available geospatial linked datasets use this CRS. %, and then, the interlinking process of GEOFLA with other web datasets will be facilitated.
For identifying resources, a base URI is constructed (according to W3C best practice) as the following: \lstinline[breaklines=true,language=turtle]{<http://data.ign.fr/id/geofla/[types_of_administrative_division]/[id]>}. For example the commune of "Saint-Mand\'e" will be identified by: \lstinline[breaklines=true,language=turtle]{<http://data.ign.fr/id/geofla/commune/94067>}.\\
In order to illustrate the fact that geometries can be described with different CRSs, the points representing respectively the location of the head office and the location of the centroid of administrative units are represented by Lambert-93 coordinates. See for example the location of the head office of the "Seine et Marne" department: \url{http://data.ign.fr/id/geofla/departement/Point_94028}. %Figures~\ref{fig:gui} shows the graphical user interface of the Datalift GeomRDF module conceived to perform this conversion.
%\begin{figure}[htp]
%  \centering
%  \subfloat{\includegraphics[scale=0.80]{datalift1}}
%  \hspace{5pt}
%  \subfloat{\includegraphics[scale=0.80]{datalift2}}
%  \caption{GUI of the Datalift GeomRDF module}
%  \label{fig:gui}
%\end{figure}
%\vspace{-0.38cm}
\subsubsection{Mapping RDF data to vocabularies} An ontology describing the GEOFLA dataset was developed and published by the IGN. Two modules of the Datalift platform can be used to map  the RDF data, generated in the previous step, to the ontology classes and properties: "RDF conversion based on ontologies" or "RDF2RDF transformation". We have used the latter which performs SPARQL CONSTRUCT queries in order to generate a new data named graph according to a given pattern. Listing~\ref{ttl1}  (section 3) shows an example of RDF triples transforming according to the GEOFLA ontology.
%\vspace{-0.38cm}
\subsubsection{Publication of the datasets} In this step, the RDF data are published online as Linked Open Data and accessible through the following SPARQL endpoint address: \url{http://data.ign.fr/id/sparql}.
%\vspace{-0.38cm}
\subsubsection{Interlinking with other datasets} As shown in the listing~\ref{ttl2}, we have used the "RDF2RDF transformation" module and administratives units identifiers in order to add owl:sameAs links with the INSEE (French National Institute for Statistics and Economic Research) web dataset. A data interlinking module based on the Silk Framework \cite{volz09a} is also available within the Datalift platform when more complicated interlinking processes are needed.
\begin{lstlisting}[basicstyle=\scriptsize\ttfamily,breaklines=true,language=turtle,caption=sameAs links between the administrative unit of Paris in IGN and INSEE, label={ttl2}]
@prefix owl: <http://www.w3.org/2002/07/owl#>

<http://data.ign.fr/id/geofla/arrondissement/751> owl:sameAs <http://id.insee.fr/geo/arrondissement/751>
<http://data.ign.fr/id/geofla/departement/75> owl:sameAs <http://id.insee.fr/geo/departement/75>
\end{lstlisting}
\vspace{-0.5cm}
\subsubsection{Spatial queries over structured geometries} As explained in section 2, we have decided to generate structured geometries in order to enable users perform basic spatial queries in SPARQL, even without storing our geospatial data in a triplestore that implements GeoSPARQL standard. The following query illustrates that scenario. It returns all the instances of rdf:Type geofla:Departement whose geometries intersect a bounding box whose coordinates are given as parameters of the query.

\begin{lstlisting}[basicstyle=\scriptsize\ttfamily,breaklines=true,language=turtle,caption=Example of a spatial query over structured geometries in SPARQL, label={ttl2}]
PREFIX rdf:<http://www.w3.org/1999/02/22-rdf-syntax-ns#>
PREFIX rdfs:<http://www.w3.org/2000/01/rdf-schema#>
PREFIX geofla:<http://data.ign.fr/def/geofla#>
PREFIX geom:<http://data.ign.fr/def/geometrie#>

SELECT DISTINCT ?name WHERE {
?dep rdf:type geofla:Departement .
?dep rdfs:label ?name .
?dep geom:geometry/geom:polygonMember/geom:exterior/geom:points ?pl .
?pl rdf:type geom:PointsList .
?pl (rdf:rest*/rdf:first)|geom:firstAndLast ?pm .
?pm rdf:type geom:Point .
?pm geom:coordX ?x .
?pm geom:coordY ?y .
FILTER ((?x > 1 && ?x < 3) &&(?y > 42 && ?y < 44)) .
}
\end{lstlisting}

This query can be applied directly on GEOFLA dataset thanks to this SPARQL Endpoint \url{http://data.ign.fr/id/sparql}. Results are the departments called "Ari\`ege", "Tarn", "Tarn-et-Garonne", "Aude", "Aveyron", "Pyr\'en\'ees-Orientales", "Haute-Garonne", "Gers" and "H\'erault".

\section{Conclusion and Future Work}
In this paper we presented the GeomRDF tools that enables the transformation of geospatial data into RDF. We presented the input format that GeomRDF supports, its mains components, and its dependencies on external libraries. We detailed how the geometric informations are structured according to an ontology that re-uses and extends the existing geographic vocabularies, and showed an example of this structuration in the case of a Multipolygon geometry.\\
The GeomRDF tool has been implemented as a module in the Datalift platform. We illustrated an application of this tool by detailing the process of publishing the French Administrative Units dataset provided by IGN France.\\

So far GeomRDF takes only geospatial formats as input data. However, it would be useful to make it also accept text file formats, such as CSV, containing direct location information described by coordinates, and convert these textual coordinates into instances of the class geom:Point. Then, such location information could be handled by query, interlinking or visualization tools in the same way as any other piece of geospatial information.

%It will be devoted also to add a functionality that handles and transforms geometric information that are included in a huge number of datasets, published in the context of the Open Government Data initiatives (e.g., \url{http://www.data.gouv.fr}), and available in formats different from the usual GIS data model (e.g., CSV, XLS, XML).

\section{Acknowledgements}
This work has been partially supported by the French National Research Agency (ANR) within the Datalift Project, under grant number ANR-10-CORD-009.

%\begin{thebibliography}{4}
%[15] PanagiotisA.Vretanos.Webfeatureserviceimplementationspecification.1.1.0(May),2005.
%[7] S. Hellmann, J. Unbehauen, A. Zaveri, J. Lehmann, S. Auer, S. Tramp, H. Williams, O. Erling, T. T. Jr., K. Idehen, A. Blumauer, and H. Nagy. Re- port on knowledge extraction from structured sources. Technical Report LOD2 D3.1.1, 2011. http://lod2.eu/Deliverable/D3.1.1.html.

%[3] S?oren Auer, Sebastian Dietzold, Jens Lehmann, Sebastian Hellmann, and David Aumueller. Triplify: Light-weight linked data publication from relational databases. In Juan Quemada, Gonzalo Le?on, Yo?elle S. Maarek, and Wolfgang Nejdl, editors, Proceedings of the 18th Interna- tional Conference on World Wide Web, WWW 2009, Madrid, Spain, April 20-24, 2009, pages 621?630. ACM, 2009.
%\end{thebibliography}
%\raggedright

\bibliographystyle{splncs03}
\bibliography{GeoConverter_VF}

\end{document}